# A blind zone-suppressed hybrid beam steering for solid-state Lidar

CHAO LI, XIANYI CAO, KAN WU*, GAOFENG QIU, MINGLU CAI, GUANGJIN ZHANG, XINWAN LI, AND JIANPING CHEN

*State Key Laboratory of Advanced Optical Communication Systems and Networks, Department of Electronic Engineering, Shanghai Jiao Tong University, Shanghai 200240, China*
*Corresponding author: kanwu@sjtu.edu.cn*

**We demonstrate a blind zone-suppressed and flash-emitting solid-state Lidar based on lens-assisted beam steering (LABS) technology. As a proof-of-concept demonstration, with a design of subwavelength-gap one-dimensional (1D) long-emitter array and multi-wavelength flash beam emitting, the device was measured to have 5%-blind zone suppression, 0.06°/point-deflection step and 4.2 μs-scanning speed. In time-of-flight (TOF) ranging experiments, Lidar systems have field of view of 11.3°×8.1° (normal device) or 0.9°×8.1° (blind-zone suppressed device), far-field number of resolved points of 192 and a detection distance of 10 m. This work demonstrates the possibility that a new integrated beam-steering technology can be implemented in a Lidar without sacrificing other performance.**

## 1. INTRODUCTION

Light detection and ranging (Lidar) technologies have been widely applied in autonomous driving, sensing, wind detection, etc. In recent years, all solid-state Lidar systems have attracted wide attention for their high potential to achieve a revolutionary performance. The key point is to replace the bulky mechanical beam-steering components with the solid-state non-mechanical beam-steering components. Various non-mechanical beam-steering technologies have been proposed including lens-assisted beam steering (LABS) [1-10], optical phased array (OPA) [11-22], micro-electro-mechanical systems (MEMS) mirror [23-25] and liquid crystal [26, 27], etc. However, fundamental limits emerge when they solve the problem of non-mechanical beam steering. For example, OPA-based beam steering has relatively high control complexity, requirement of precise analog control and limited sidelobe/background suppression. MEMS mirror-based beam steering has a limited steering speed as well as the potential problem of mechanical fatigue. In recent years, LABS has attracted more and more attention for its advantages of very low control complexity and high sidelobe/background suppression. It consists of a 1×$N$ switch, $N$ emitters connected to $N$ outputs of the switch and a lens. The light beam is guided to one of the emitters through the switch and emits to the free space. Then the lens collimates the beam and steers its direction. If the 1×$N$ switch is realized by a binary tree structure of 1×2 switches, only $\log_2 N$ 1×2 switches work simultaneously, corresponding to a control complexity and power consumption of O($\log_2 N$). Meanwhile, because only one emitter emits light beam in any time, a sidelobe/background suppression more than 20 dB can be easily obtained.

Various LABS devices have been explored with different features including ring switches [5], integrated planar lens [1, 3], binary switch tree [6, 9], metalens [4], photonic crystal waveguide (PCW) grating [2, 8, 28, 29], MEMS emitter [10] and MEMS optical switch [7]. However, limitations also exist in LABS technology such as blind zone and steering speed, which come from the beam-steering principle and limited performance of basic components e.g., lens, emitter and optical switch.

The existence of blind zone in the field of view (FOV) results from the principle of discrete beam steering in LABS devices. As the far-field FOV is exactly the image of the beam pattern on the emitter plane, any gap between two emitters will lead to a blind zone in the FOV. Therefore, blind zone can be suppressed by increasing the fill factor of the emitter array, i.e., the emitters should be placed as dense as possible. For 2D emitter array, the emitter fill factor is limited by switch size, waveguide-bending radius and/or emitter size. In a few reported architectures with either 2D ring emitter array [4], MEMS emitter array [10] or normal grating emitter array [5-7], the fill factors are typically only 5% or less. A hybrid architecture with 1D long emitter array and wavelength tuning can improve the fill factor as the emitters can be placed close to each other. 1D PCW grating array [2, 8, 28, 29] has been proposed to enhance the angular dispersion for wavelength-assisted beam steering as well as to improve the fill factor. But the propagation loss of PCW is higher than strip waveguide.

The steering speed of LABS device is another issue as it determines the target detection speed in Lidar applications. The widely used thermo-optic switches [1-6, 8, 9, 30] typically have 10s μs switching speed, which is slow. Electro-optic switches in silicon [31, 32] can achieve nanosecond-level speed, but have higher loss and are less capable of handling high optical power. MEMS based switches [7, 10] have few microsecond speed, but can be potentially vulnerable to vibration and mechanical fatigue. Therefore, a more advanced Lidar design must be employed to overcome the speed limitation in LABS technology.

In this work, we demonstrate a hybrid LABS-based Lidar system that overcomes the fundamental limits in LABS. Firstly, an off-chip cylindrical lens is setting on switchable 1D long-emitter array to obtain high beam quality. Then with a novel Lidar structure that implements a parallel flash beam emitting and subwavelength-gap 1D long-emitter array, the blind zone, one fundamental limit of LABS, is significantly reduced to 5%, corresponding to a deflection angle resolution of 0.06°/point. Moreover, the detection speed, another fundamental limit, can be multiplied by the number of parallel channels. In the proof-of-concept experiment, the Lidar works under time-of-light ranging mode with 16×12-point beam steering, 11.3°×8.1° field of view and 25-dB background suppression. A blind-zone suppressed LABS Lidar with a FOV of 0.9°×8.1° is also implemented for ranging experiment. This work demonstrates that the fundamental limits of LABS can be overcome with proper Lidar design without sacrificing other performance, which would be a solid step towards a truly practical all solid-state Lidar for various applications including autonomous driving and 3D imaging, etc.

## 2. PRINCIPLE OF 2D BEAM STEERING

To overcome the fundamental limits of blind zone and limited beam-steering speed, a wavelength-assisted LABS design is employed, as shown in Fig. 1. The device consists of an on-chip 1×16 switch with thermal control and 1D emitter array with staircase structure and an off-chip cylindrical lens. The light source is coupled into the chip by a standard lensed fiber. The silicon nitride waveguide was designed with a size of 1 μm × 0.4 μm to support TE/TM mode operation near 1550 nm. The 1×16 switch is realized by a binary tree of cascaded 1×2 Mach-Zehnder interferometer (MZI) switches with thermal control, where the heating resistors are titanium film deposited on silica cladding, and the wiring lines and bonding pads are aluminum. A cross section of the switch is depicted in the inset (i) of Fig. 1. The 1D emitter array consists of 16 grating emitters. And 16 emitters are connected by 16 outputs of the switch. The grating emitter was design to have a staircase structure with three stages as shown in the inset (ii) of Fig. 1. Above the chip, an off-chip cylindrical lens is set and its focal plane is overlapped with the plane of emitter array. Light emitted from a certain grating is collimated and steered in xoz plane. In yoz plane, the emission angle of light is controlled by the wavelength of light source and not affected by the cylindrical lens.

The beam-steering principle in two dimensions is further illustrated in Fig. 2. A circular lens denoted as "FT lens" is set in the setup to perform the Fourier transform and obtain the far-field beam pattern of the light emitted from the system. The FT lens has the same focal length with the cylindrical lens. In xoz plane, as shown in Fig. 2(a), beam steering is based on LABS technology. Light beams emitted from different grating emitters are collimated and steered by the cylindrical lens. It should be emphasized that these light beams do not exist at the same time because each time only one grating emitter is turned on to emit light. But these light beams virtually intersect at one point (denoted as "S") on the focal plane of the cylindrical lens on the other side. This means by switching light to emit from different emitters, different directions of beam steering in the far field can be achieved. The beam divergence $\Delta\theta_{xoz}$ and FOV $\theta_{xoz}$ in xoz plane are given by [6]:

$$\Delta\theta_{xoz} = tan^{-1}(w/f) \tag{1}$$

$$\theta_{xoz} = 2tan^{-1}(l/2f) \tag{2}$$

where $w$ is the near-field beam diameter, $l$ is the size of the emitter array and $f$ is the focal length of two lenses. The beam steering step $\theta'_{xoz}$ can be expressed as:

$$\theta'_{xoz} \approx tan^{-1}(p/f) \approx \theta_{xoz}/(N-1) \tag{3}$$

where $p$ is the distance between two adjacent emitters, $N$ is the number of emitters. As each emitter corresponds to a beam direction in the far field, $N$ emitters correspond to $N$ different beam directions and thus the number of resolved points in the far field is $N$. In yoz plane, as depicted in Fig. 2(b), light beams of different wavelengths emitted from the same grating diffract into different directions according to the grating diffraction equation given by [33]:

$$\Phi = sin^{-1}(\beta/k_0 - \lambda/\Lambda) \tag{4}$$

where $\Phi$ is the diffraction angle, $\lambda$ is the wavelength, $\beta$ is the propagation constant of light mode in the grating, $k_0$ is the wave number in vacuum, $\Lambda$ is the grating period. The FOV angle $\varphi_{yoz}$ can be derived by differentiating Eq. (4) with respect to $\lambda$ as [28]:

$$\varphi_{yoz} = -\frac{n_g - n_r}{\lambda\sqrt{1-n_r^2}}\Delta\lambda \tag{5}$$

where $n_g = -(\lambda^2/2\pi)\cdot(d\beta/d\lambda)$ is the group index, $n_r = (\beta/k_0-\lambda/\Lambda)$ is the modal equivalent index of the radiated light, and $\Delta\lambda$ is the wavelength range of input light. The beam divergence $\Delta\varphi_{yoz}$ is expressed as follows based on Fraunhofer diffraction theory [33].

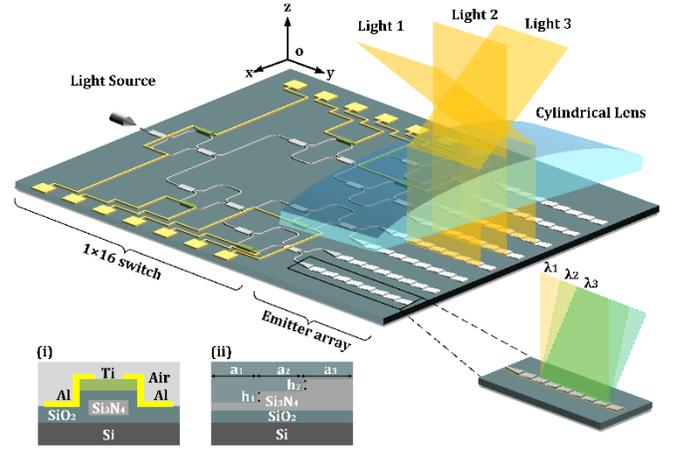

**Fig. 1.** Schematic illustration of cylindrical lens-based beam-steering device. Inset: cross-section of (i) thermal switch and (ii) staircase grating.

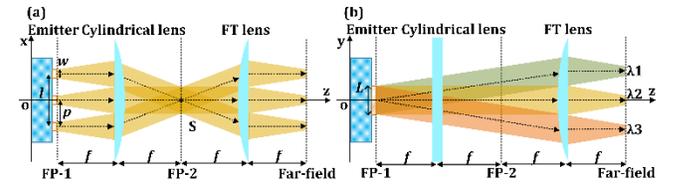

**Fig. 2.** Principle of beam steering in (a) xoz plane, and (b) yoz plane.

$$\Delta\varphi_{yoz} = 0.886 \times \lambda/L \tag{6}$$

where $L$ is the effective grating length when light power decreases to $1/e$. The divergence and emitting angles remain unchanged in yoz plane when light beams propagate through cylindrical lens. Similarly, the far-field beam pattern in yoz plane is then obtained by the FT lens. According to Eqs. (4)-(6), both beam divergence and diffraction angle in yoz plane depend on the wavelength of light. And the FOV in yoz plane is determined by the wavelength range.

To briefly summarize, beam steering in the xoz plane is determined by the beam pattern on the emitter plane. The blind zone in the far field is generated by the gap between two adjacent grating emitters. Beam steering in the yoz plane is determined by the wavelength tuning, which is gapless in the far field and does not create blind zone. Moreover, simultaneously emitting multiple wavelength (flash beam emitting) can multiply the detection speed by the number of wavelength channels. Although similar hybrid architectures are also proposed in some LABS devices, the reported works all have limited performance in some key specifications. A detailed performances comparison among our work and other hybrid architectures has been listed in Table 1. It is clear that our work exhibits a balanced performance in nearly all key specifications.

**Table 1. Performance comparison of hybrid LABS device**

| Ref | Structure | Beam quality | Blind zone | Main loss (value) | Scanning method |
|---|---|---|---|---|---|
| [1] | Planar lens+1D grating | Aberration | >50% | a-Si lens (NA) | Thermo-optic switch + wavelength tuning |
| [8] | Prism lens+1D PCW grating | NA | 87.5% | PCW (10~30 dB/cm) | Thermo-optic switch + wavelength tuning |
| Our work | Cylindrical lens+1D grating | 25 dB background suppression | 5% | Grating (1~3 dB/cm) | Thermo-optic switch + flash emitting |

## 3. DEVICE CHARACTERIZATION

### A. Device fabrication

The chip was fabricated on a 6×16-mm$^2$ silicon-nitride on isolator wafer with a silicon photonic CMOS process. A minimum feature size of 200 nm in grating structure is achieved by electron beam lithography (EBL) process. Figure 3(a) shows the photograph of the chip. It includes 4 stages of cascaded 1×2 MZI switches and 16 grating emitters placed in parallel. The inset picture in Fig. 3(a) illustrates the radiation beam pattern of the staircase grating captured by an infrared camera (Xenics Bobcat-320). Figure 3(b) shows the setup of our beam-steering device, where a cylindrical lens with a focal length of 10 mm (Thorlabs LJ1878L2-C) is set above the chip. The thermo-optical switch on chip was measured to have a rise time (10%-90%) of ~50 μs, as shown in Fig. 3(c), and of which the power consumption and extinction ratio were measured to be ~120 mW/π and ~25 dB.

Generally, the emission efficiency of the Si3N4 Bragg grating is limited because of the low refractive index contrast compared with silicon. Methods including distributed Bragg reflector [34], staircase structure [35], dual-level grating structure [36] are proposed to increase the directionality of Si3N4 grating. Among these structures, the staircase grating is demonstrated to support an efficiency of 71% with single layer structure. Besides, an enhanced upward emission efficiency of 50%~90% can be obtained for a staircase grating with a radiation length of 40 μm in the FDTD simulation whereas a normal grating with single full-depth etching can only have an efficiency below 50%. The high directionality results from the constructive interference in upward direction between scattering lights from two consecutive trenches. Therefore, in our device, as shown in Fig. 3(d), the emitters were fabricated as staircase gratings to increase the directionality. Considering that a grating with long radiation length is required to obtain small beam divergence in the far field according to Eq. (6), there is a tradeoff between the grating length and emission efficiency during the grating design. A parameter optimization based on particle swarm algorithm is applied in simulation. As shown in the inset of Fig. 3(d), the staircase grating was fabricated with a width of ~11 μm, a length of ~1 mm, an etching depth of 50 nm and a period of ~0.96 μm. The upward emission efficiency is simulated to be 64%. Experimental emission efficiency of long waveguide grating is obtained by light beam brightness comparison with standard grating coupler in infrared image.

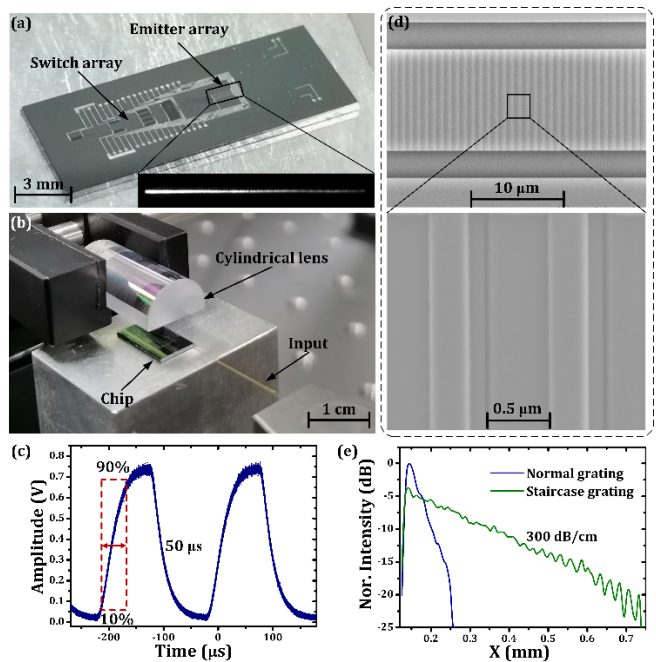

**Fig. 3.** (a) Photograph of the chip with switch and emitter array. Inset: Radiation intensity of staircase grating captured by an infrared camera. (b) Photograph of the beam-steering device. (c) Speed measurement of the thermo-optical switch. (d) SEM image of the staircase grating. Inset: zoom-in image. (e) Comparison of radiation intensity between staircase and normal gratings.

As shown in Fig. 3(e), the staircase grating was measured to radiate 23% more power than a normal single-etched grating (50%). It has an attenuation coefficient of 300 dB/cm along the propagation direction, and an effective grating length of 145 μm (power decreases to 1/e). The spacing between two adjacent gratings is designed with two values of 121 μm and 0.6 μm. The device with large grating spacing is to achieve a relatively large FOV with limited grating number according to Eq. (2). The device with small grating spacing is to demonstrate that the blind zone of device with 121 μm-spacing can be significantly suppressed according to Eq. (3).

### B. Two-dimensional beam steering

2D non-mechanical beam steering in our device is achieved by guiding the light to emit from a certain emitter in the array and tuning the input wavelength. In the experiment, a wavelength-tunable continuous wave laser is coupled into the chip through a lensed fiber for wavelength-assisted beam steering demonstration. The light is routed into one of the 16 emitters by the cascaded 1×2 MZI switches on the chip. Above the chip, a cylindrical lens is utilized to collimate and steer the up-emitting light in xoz plane, as described in section 2.

We first characterize the grating emission properties by removing the cylindrical lens and placing a Fourier transform lens (FT lens, a circular plano-convex lens with 10-mm focal length), as shown in Fig. 4(a). The output of the FT lens is captured by an infrared camera with a detection dynamic range of 0~65535. As shown in Fig. 4(b), the measured far-field grating emission pattern exhibits a wide divergence angle (~8.9°) along θ direction due to the narrow grating width and a narrow divergence angle (~0.5°) along φ direction due to the long grating length. When the wavelength changes from 1455 nm to 1565 nm, the emission angle shifts ~8.1° along φ direction. In Fig. 4(b), a few typical measured patterns with different input wavelengths are plotted together for a clear view. Then the cylindrical lens is placed together with the FT lens for the far-field measurement of the beam-steering device (grating spacing of 121 μm), as shown in Fig. 4(c). Similarly, all the measured far-field patterns are plotted together in Fig. 4(d) for a clear view. The FOV is ~11.3° along θ direction based on LABS and ~8.1° along φ direction based on wavelength tuning. This confirms the property that addition of cylindrical lens does not affect the beam-steering angle based on wavelength tuning. A nearly linear relation of ~0.07°/nm between wavelength and grating emission angle (φ) is obtained in both simulation and experiment. Based on LABS principle, there is only one grating emitting light each time, and the background noise is dominated by the leaked power from un-working emitters through non-ideal switches. Therefore, the background suppression is determined by the extinction ratio of switches (upper panel of Fig. 4(e)), which is 25 dB. Except for the power leakage from other emitters, more than 40 dB background suppression can be realized as depicted in the lower panel of Fig. 4(e), which is the cross-sectional power of beam patterns along θ direction with φ = 7° in Fig. 4(d). Moreover, the typical cross-sectional powers of the beam patterns in Figs. 4(b) and 4(d) are shown in Fig. 4(f). It can be seen that after the collimation of cylindrical lens, the divergence of the light beam from the grating has been compressed from 8.9°×0.5° in θ and φ direction (Fig. 4(f) upper panel) to 0.1°×0.5° (Fig.4(f) lower panel), which is very close to the theoretical value of 0.063°×0.54°.

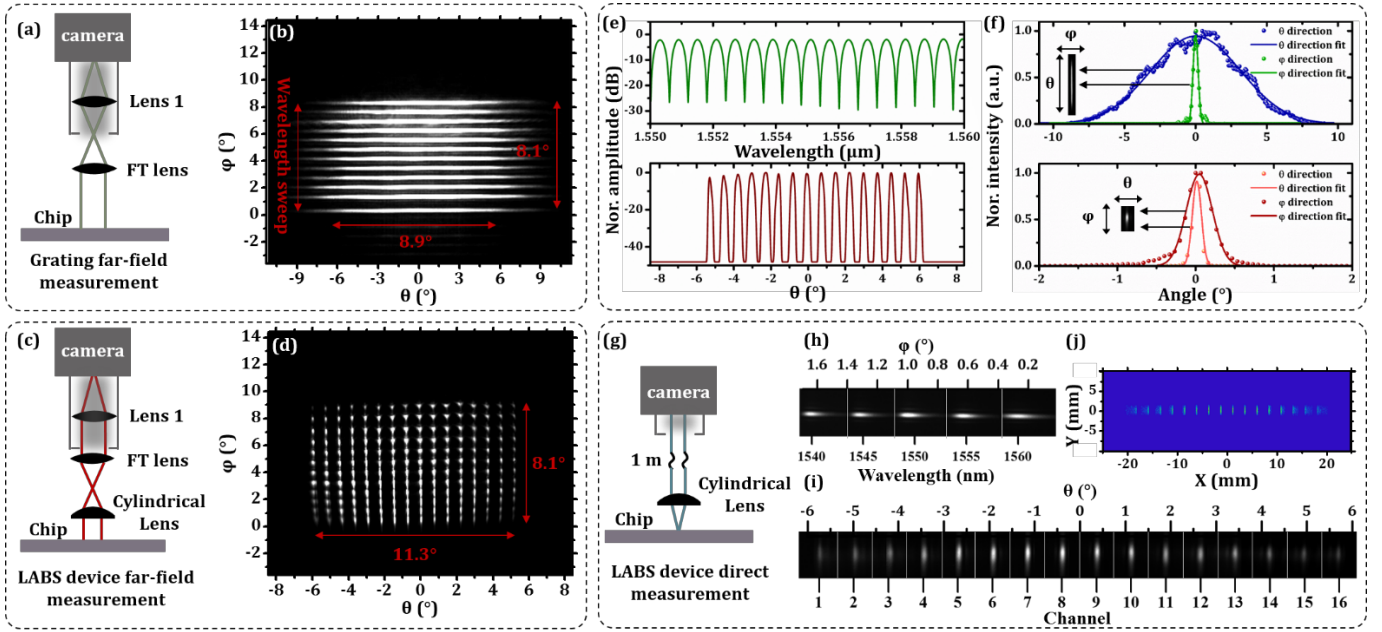

**Fig. 4.** (a) Experimental setup for far-field measurement of grating. (b) Far-field beam pattern of grating with wavelength range of 110 nm. (c) Experimental setup for far-field measurement of LABS device. (d) Far-field beam-steering pattern by emitter selecting and wavelength tuning. (e) A typical transmission spectrum of a 1×2 MZI switch with unequal arms (upper panel) and cross-sectional powers of beam pattern along θ direction in 1550 nm (lower panel). (f) Cross-sectional powers and fitted results of far-field beam pattern from grating (upper panel) and LABS device (lower panel). Inset: single beam patterns from b) and d). (g) Experimental setup for direct measurement of far-field beam patterns from LABS device. (h) Beam-steering patterns by wavelength tuning. (i) Beam-steering patterns by LABS. (j) Beam-steering patterns by LABS in Zemax simulation.

To further confirm the measurement accuracy of the steering angles, we also measured the far-field beam patterns by directly placing the infrared camera 1 m away from the device, as shown in Fig. 4(g). Different far-field angles were measured by moving the camera. The results are shown in Figs 4(h) and 4(i) from wavelength tuning and LABS, respectively. Similarly, all the measured beam patterns are plotted together in the figures. The steering angle is ±5.65° for LABS in θ direction and 1.47° for wavelength tuning (1540~1560 nm) in φ direction. Both agree well with the measurement results based on FT lens. There is an intensity reduction of beam pattern from center emitter to edge emitter, which is a result of aberration from the cylindrical lens. The same result is confirmed by simulation, as shown in Fig. 4(j). Generally, it can be improved by utilizing a better lens system.

### C. Blind zone suppression

As explained in Section 2, 1D long-emitter array in our design allows densely placed grating emitters and the gap between adjacent gratings can be significantly reduced, corresponding to a highly suppressed blind zone in the far field. Our previous work shows a blind zone suppression with defocusing method but the cost is the increase of beam divergence. [6] To achieve a small gap, one needs to reduce the crosstalk between two adjacent grating emitters. In this work, a grating with large core area is designed so that the light beam can be well confined in the waveguide. For a cross section of 11×0.4 μm² and a grating length of 2.5 mm, Figure 5 shows the experimental and simulated crosstalk with different grating gaps. Crosstalk between two adjacent gratings is simulated to be -60 dB and -30 dB with gaps of 1.5 μm and 0.6 μm. Experimental measurement of crosstalk between gratings has also been implemented. We fabricated two waveguide grating emitter arrays with emitter gaps of 1.5 μm and 0.6 μm and an adequate coupling length of 2.5 mm. When light is guided into one grating of emitter array, the crosstalk is -48 dB and -17 dB respectively. The micrographs are shown in Figs. 6(a) and 6(b), respectively. Besides, further evidence such as dark lines between each beam pattern can be observed in the zoomed-in images of two types of emitter arrays when all the gratings are lit up as displayed in Figs. 6(c) and 6(d).

Further demonstration of blind-zone suppressed beam steering is carried on our device with 0.6-μm-gap waveguide grating emitter array. We have obtained a far-field FOV of 0.91° based on LABS and an FOV of 8.1° based on wavelength tuning in our device, as depicted in Fig. 7(a). Both vertical and horizontal beam steering are shown individually in Figs. 7(b) and 7(c) for a clear observation. It has a same FOV by wavelength tuning as we have proved in Figs. 4(a) and 4(b), but a higher resolution point by LABS in the same FOV. For example, there are only two points in an FOV of 0.91° in a 121-µm-gap grating array, and 16 points in the same FOV in a 0.6-µm-gap grating array, the minimum deflection angle has been decreased from 0.75°/point to 0.06°/point. As shown in Fig. 7(c), there is no blind zone observed between two adjacent channels of beam pattern. Though the FOV of device with 0.6-µm-gap grating array decreased under the same number of emitters, it can be enhanced to 11.3 ° by fabricating 170 close-packed waveguide gratings.

If we define a parameter "blind zone suppression (BZS)" as the ratio between the total emission beam area and the total area occupied by the emitters, our design supports an BZS of 88% (=11 µm /(11+1.5) µm) and a maximum BZS of 95% (=11 µm /(11+0.6) µm) in experiment, that is, the blind zone only occupies 5% of the FOV. Other reported LABS Lidar systems all have a blind zone more than 50% (see Discussion).

## 4. LIDAR DEMONSTRATION

The Lidar that overcomes the fundamental limits of LABS is then demonstrated, as shown in Fig. 8. The ranging technology is time-of-light (TOF) ranging. A femtosecond pulsed laser with a broadband spectrum near 1550 nm is utilized as the light source. After the laser, a pulse picker is applied to reduce the laser repetition rate from 37 MHz to 20 kHz. Then a spectral filter is introduced to select the desired spectrum as different wavelength channels. The schematic illustrations of the corresponding spectra and waveforms are shown as insets (i)-(iii) in Fig. 8. The pre-processed pulsed light source is divided into two paths. 10% of it is guided to a photodetector as timing reference. 90% of it is amplified by an Erbium-doped fiber amplifier (EDFA), controlled by a polarization controlled (PC) and coupled into the chip via a lensed fiber. On the chip, the light is guided to one certain grating emitter by the 1×16 switch and emits to the free space. The pulses with different wavelengths overlap in the time domain. So they simultaneously emit to different directions from the grating emitter (flash beam emitting). After the collimation of the cylindrical lens, the collimated beams simultaneously detect the target along φ direction (wavelength dependent direction). The beams are then scanned along θ direction (LABS direction) by switching the light to different grating emitters on the chip. The reflected beams are collected by a receiver consisted of a lens, a fiber array and a few avalanche photodiodes (APDs). In principle, fibers receiving the reflected beams along θ direction (LABS direction) do not work in the same time, so their output can be detected by a same APD. Therefore, the required number of APDs is equal to the number of wavelength channels. In the future, a 1D APD array can be used. The electrical signals output by the APDs together with the reference signal are finally processed to obtain the timing difference and target distance. A photograph of the experimental setup is shown in Fig. 9.

The advantage of such a Lidar design is that it fully exploits the low control complexity and clean beam emission of LABS while overcoming its fundamental limits of blind zone and relatively low speed. The blind zone is eliminated in y direction by wavelength tuning and suppressed in x direction by reducing the gap between emitters. The speed is improved by flash beam emitting. That is, all the light pulses with different wavelengths emits simultaneously. Such a parallel detection multiplies the detection speed per pixel by the number of wavelength channels. The experimentally measured spectra and time-domain waveforms of the pre-processed light source are shown in Figs. 10(a) and 10(b). In this proof-of-concept demonstration, twelve wavelength channels from 1544 nm to 1562 nm are selected. The channel spacing is 1.6 nm, corresponding to an emission angle difference of 0.1°. The repetition rate is 20 kHz. The pulse duration is ~500 ps. The loss of our chip is about ~-12 dB including -4 dB of fiber-to-chip coupling loss, -6 dB of 1×16 switch insertion loss (1.5 dB each 1×2 MZI switch), and -2 dB of grating emission loss. The average output power of our device is measured to be -30.51 dBm. The target is a white A4 paper. The power of returning light Pr can be expressed as:

$$P_r = \frac{P_t A_r^2 r_t t_a}{8R^2}$$

(7)

where $P_t$ is the power of emitted light from the transmitter, $A_r$ is the receiver aperture, $r_t$ is the reflectivity of target, $t_a$ is the air transmission and $R$ is the target distance. The receiver lens has a diameter of 2.05 cm and an amplified APD with a response bandwidth of 200 MHz is used. The target distance can be obtained by calculating the time delay between the received signal and the reference signal. Fig. 10(b) shows two typical returning signals at a distance of 1.08 m and 11.22 m, respectively. The loss related to the beam propagation and target diffuse reflection is estimated to be ~50 dB according to Eq. (7). The target at 11.22 m is a high reflective mirror for proof-of-concept demonstration.

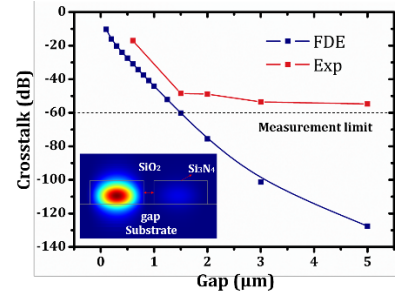

**Fig. 5.** Crosstalk between two grating emitters with different gaps. Inset: simulated mode distribution from a finite-difference-eigenmode (FDE) solver.

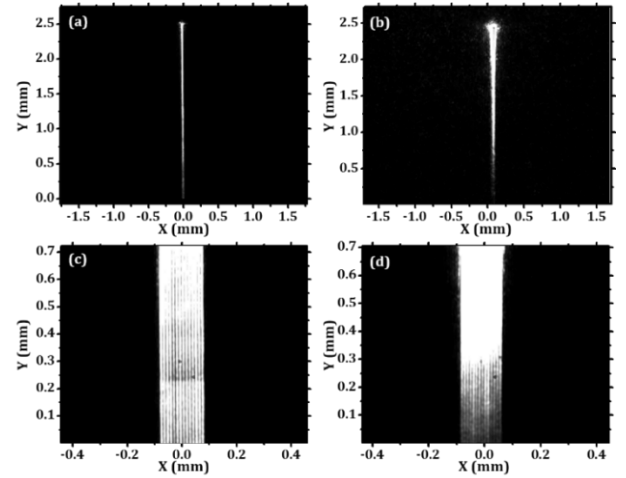

**Fig. 6.** Near-field beam pattern with one grating lit up in grating emitter array with gaps of (a) 1.5 µm, and (b) 0.6 µm. Near-field beam pattern with all gratings lit up with gaps of (c) 1.5 µm, and (d) 0.6 µm.

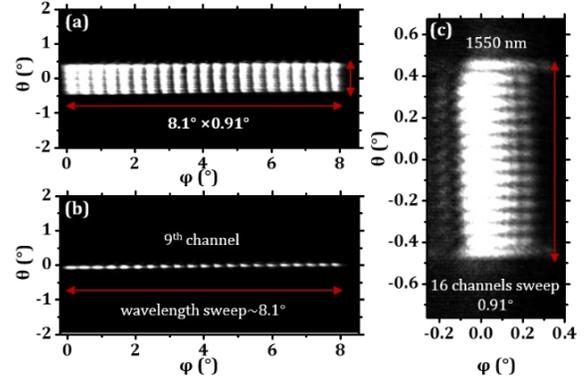

**Fig. 7.** Far-field beam-steering pattern realized by (a) 16-channel 0.6-µm-gap emitters selecting and wavelength tuning, (b) wavelength sweep, and (c) 16-channel emitters selecting.

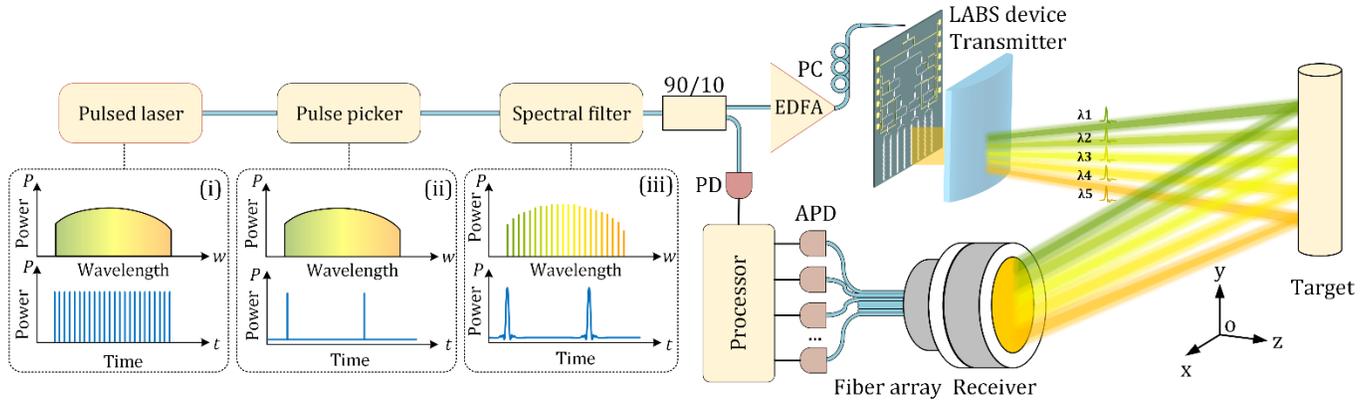

**Fig. 8.** Experimental setup of target detection with the beam-steering device. Inset: the output spectra and waveforms of (i) pulsed laser, (ii) pulse picker and (iii) spectral filter

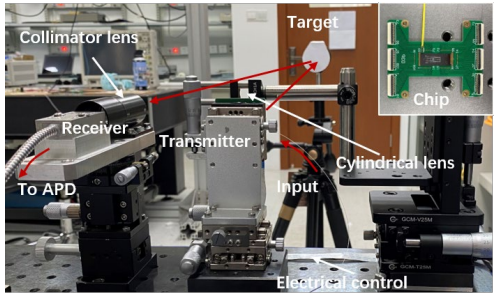

**Fig. 9.** Picture of experimental setup of Lidar. Inset: picture of LABS chip with electrical control.

To evaluate the ranging accuracy, three groups of ranging experiments by LABS (measurement 1, 2) and wavelength tuning (measurement 3) are performed. Targets are placed with a separation of 15 cm within a range of 3 m in the horizontal direction (LABS direction) and a separation of 20 cm within a range of 1.5 m in the vertical direction (wavelength direction), respectively. The corresponding waveforms of the returning signals are provided in Fig. 11. The comparison of measurement results between TOF ranging and manual measurement are shown in Fig. 12(a). The maximum deviation are 0.35 ns corresponding to 5 cm ranging difference, which is mainly due to the manual measurement error. The theoretical ranging error $\Delta R$ of TOF lidar is given by [9]:

$$\Delta R = 0.5(\frac{\tau_r}{SNR})c \qquad (8)$$

where $\tau_r$ is the pulse rise time, $SNR$ is the signal-to-noise ratio of the returning signal and $c$ is the speed of light in vacuum. For our system, the pulse rise time is 0.825 ns limited by the bandwidth of amplified APD (200 MHz), the SNR of the returning signal is ~46.7. Therefore, the ranging error is calculated to be ~2.65 mm.

Additionally, a proof-of-concept demonstration of blind-zone suppression in Lidar system has been performed. A comparison experiment of ranging between devices with 121-μm-spacing grating array and 0.6-μm-spacing grating array. The ranging results along emitter direction are shown in Fig. 12(b). Though the beam-steering device with 121-μm-spacing gratings has a FOV of 11.3°, the targets can be detected only when they locate in the yellow region (sight zone) in Fig. 12(b), which corresponds to a blind zone of 91.7%. However, the blind zone can be suppressed to 5% by the device with 0.6-μm-spacing grating array, as depicted in the inset of Fig. 12(b). Besides, as discussed in Section 3C, the FOV can be simply increased by increasing the number of gratings with large-scale integration.

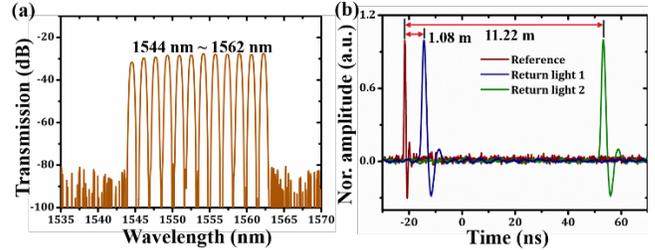

**Fig. 10.** (a) Optical spectrum of the light source. (b) Waveforms of reference signal and two returning signals scattered by the targets at 1.08 m an 11.22 m.

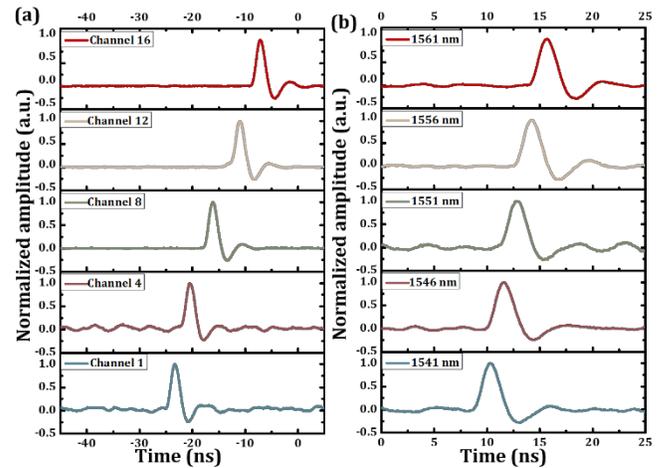

**Fig. 11.** Waveforms of returning signals from different directions along (a) emitter channels of 1, 4, 8, 12, 16, and (b) wavelength channels from 1541 nm to 1561 nm.

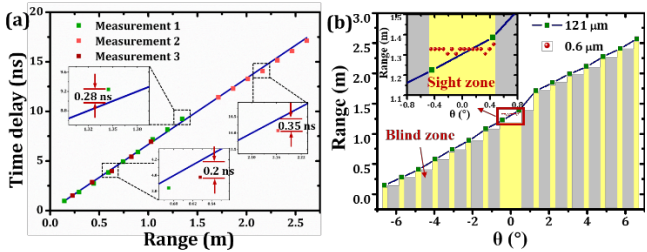

**Fig. 12.** (a) Time delays of returning pulses with respect to target distance in three measurements from 8th emitter channel at 1550 nm. (b) Ranging results comparison between beam-steering device with 121-µm and 0.6-µm-spacing grating array. Inset: zoomed view of ranging results of the device with 0.6-µm-spacing grating array.

## 5. DISCUSSION

### A. Performance comparison

Until now, although such many beam-steering technologies have been proposed, only a few Lidar demonstrations based on these new technologies [9, 10, 30, 37-41] have been reported. One reason could be that the abovementioned fundamental limits have become the biggest obstacles that prevent novel non-mechanical beam-steering technologies from practical Lidar application. A performance comparison among our work and other solid-state Lidar systems based on integrated OPA and LABS has been performed, as listed in Table 2. It can be clearly observed that only our Lidar design can support a BZS up to 95% whereas other LABS based Lidars typically have a BZS below 50%. That is, more than half of the FOV cannot be detected. Moreover, flash beam emitting design effectively counterbalances the relatively slow thermo-optic (TO) switching speed in silicon nitride. The average detection speed per pixel is 50 µs /12=4.2 µs where 12 is the number of wavelength channels. Obviously, the detection speed per pixel can be further reduced if more wavelength channels are employed. Compared with OPA-based Lidar, LABS has lower control complexity with O($\log_2 N$) complexity, digital control and better background suppression.

### B. FMCW Lidar demonstration

Recently, a growing interest in frequency-modulated continuous-wave (FMCW) ranging based on coherent detection has emerged. It is demonstrated with high sensitivity and velocity detection capability. We have also demonstrated that our LABS device can be applied in FMCW Lidar applications, as shown in Fig. 13. The light source is a distributed feedback (DFB) laser with continuous wave output. By applying a saw-like driving current to the DFB laser, a frequency-sweeping light source with a sweeping slope of 23.3 GHz/ms is realized. The sweeping light is then split into two paths. 90% of its power is amplified by an EDFA and injected into the LABS chip. The average output power is estimated to be -9.38 dBm. 10% of its power is coupled to an avalanche photodetector (APD) as local reference, together with the returning signals reflected by the target. The transmitter part is unchanged, i.e., the chip with a cylindrical lens. The receiver part is a collimation lens with a coupling fiber. Therefore, the direction of this collimation lens should be adjusted according to the beam direction in the transmitter part. As the wavelength of the DFB laser is fixed, this setup currently only supports 1D scanning along LABS direction. But an FMCW Lidar based on a microcomb with multiple wavelengths has been recently reported [42]. With such a microcomb, our setup can achieve 2D detection.

Three typical beating signals of the returning signals from 0.3 m, 0.6 m and 1 m are shown in Figs. 14(a)-14(c). The corresponding electrical spectra are shown in Fig. 14(d). Three beating frequencies fb at 341 kHz, 388 kHz and 440 kHz can be clearly observed, which correspond to an approximately linear separation in range as shown in Fig. 14(e).

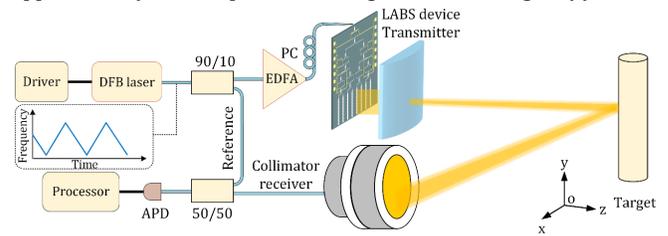

**Fig. 13.** Experiment setup of FMCW lidar system based on LABS.

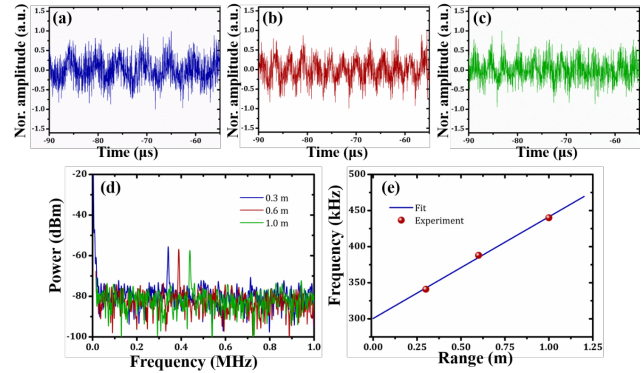

**Fig. 14.** Waveforms of beating signal from targets of (a) 0.3 m, (b) 0.6 m and (c) 1m. (d) The electrical spectra of beating frequencies from three targets. (e) Range of targets with respect to beating frequency.

### C. Future improvement of device performance

To further improve the performance of the LABS chip and Lidar system, a few factors can be considered. The first is the grating emitter. Current 2-time etching for the staircase grating increases the fabrication error, complexity and cost. A substitute method of slated grating can be introduced to obtain a high emission efficiency as well as low fabrication complexity [43]. The second is the insertion loss of the on-chip devices. It has been reported that the propagation loss in silicon nitride waveguide can be as low as ~1 dB/m with novel reflow process [44, 45]. Even with standard CMOS technology, the MZI thermo-optic switches can reach an insertion loss of ~0.6 dB with leaked power less than -25 dB [46]. So for N-channel output, the total insertion loss can be reduced to 0.6·$\log_2 N$ dB. The last is the coupling loss. With multi-stage taper design, a fiber-to-chip coupling loss of 0.5 dB can be achieved [47].

## 6. CONCLUSION

In conclusion, a solid-state TOF Lidar based on hybrid LABS technology is demonstrated on the silicon nitride platform. The implementation of 1D long-emitter array and flash beam emitting overcomes the fundamental limits of blind zone and relatively low speed in LABS technology while still maintaining its advantages of very low control complexity of O($\log_2 N$) and excellent background suppression. This work could be a very promising solution for the all solid-state integrated Lidar applications. As a proof-of-concept experiment, we have achieved 16×12-point beam steering and detection in a time-of-light ranging mode. The emission beam has 0.1°×0.5° divergence in two orthogonal directions, 11.3°×8.1° (normal device) or 0.9°×8.1° (blind-zone suppressed device) field of view and 25-dB background suppression.

**Table 2. Performance comparison of solid-state Lidars with different beam steering and ranging technologies**

| Ref | Beam steering tech. | Ranging tech. | Theoretical max BZS | Speed (μs/point) | Material | Control complexity | FOV | Dimension | Resolution |
|---|---|---|---|---|---|---|---|---|---|
| [37] | OPA with TO phase shifters | FMCW | - | 30 | Si | O(N) Analog control | 56°×15° | 2D | 200 |
| [30] | LABS with TO switches | FMCW | ~50% only for 1D | NA | Si | O($\log_2 N$) Digital control | 70° | 1D | 8 |
| [10] | LABS with MEMS switches/emitters | TOF | <10% | 4 | SiN+Si | O(1) Digital control | 1°×1° | 2D | 100 |
| [9] | LABS with TO switches | TOF | ~50% | 1000 | SiO$_2$ | O($\log_2 N$) Digital control | 1°×1° | 2D | 16 |
| This work | LABS with TO switches | TOF | ~95% /8.3% [a] | 4.2 | SiN | O($\log_2 N$) Digital control | 0.91°×8.1°/ 11.3°×8.1°[a] | 2D | 192 |

[a] The first and second numbers are from the devices with 0.6-μm and 121-μm grating spacing, respectively

**Funding.** National Natural Science Foundation of China (61875122, 61922056).

**Disclosures.** The authors declare no conflicts of interest.

**Data availability.** No data were generated or analyzed in the presented research.

# A blind zone-suppressed hybrid beam steering for solid-state Lidar: supplemental document

This document provides supplementary information to "A blind zone-suppressed hybrid beam steering for solid-state Lidar". We present the simulation analysis on 2D beam steering, design of the staircase grating emitter, evaluation of the grating upward emission efficiency, performance of silicon-nitride thermo-optical switch, and calculation of blind zone suppression.

## S1. Simulation analysis on 2D beam steering

We simulated the proposed 2D beam-steering device with FT lens in Zemax. The pitch of waveguide grating emitters is 132 μm. The cylindrical lens and FT lens form a coaxial system with the focal length of 10 mm, which is the same setup shown in Fig. 1 and Fig. 2. The near-field beam pattern of grating emitter array is shown in Fig. S1(a), which is set with a uniform intensity pattern of 11×145 μm² area for simplicity. The beam divergence and emission angle of emitter array are set as 8.9° and 0°~8.1°, respectively, which are obtained from the experiment. Figure S1(b) shows that the far-field beam pattern has a FOV of ~11.3°×8.1°, which is consistent with the analysis in Section 2. Far-field beam pattern of each emitter cannot be observed clearly in FT lens system due to the limited monitor resolution in the simulation. Instead, a direct observation of beam pattern is achieved by setting a monitor 20-cm away from the emitter array plane. As shown in Figs. S1(c)-(d), both the lens-assisted and wavelength-controlled beam steering can be clearly observed. There is an intensity reduction of beam pattern from center emitter to edge emitter, which is a result of aberration from the cylindrical lens. The same phenomena appear in the experiment as shown in Fig. 4(i). Generally, it can be improved by utilizing a better lens system.

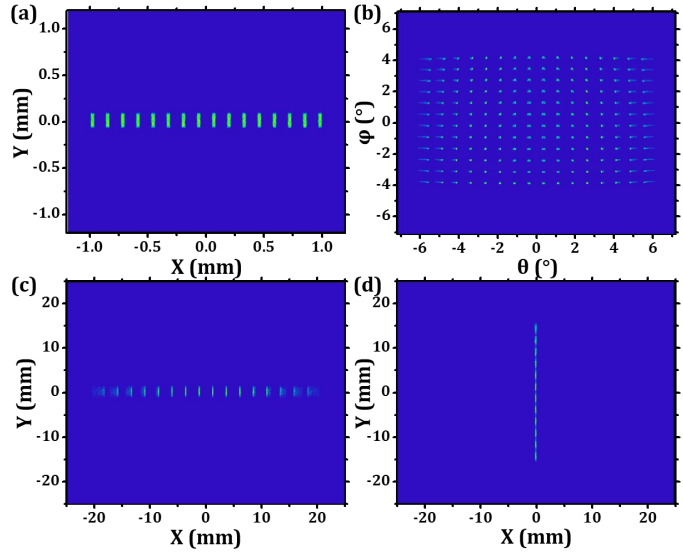

Fig. S1. Simulation results of proposed cylindrical lens-assisted beam steering. (a) Near-field beam pattern of emitter array. (b) Far-field beam pattern with all the emitters and wavelengths. (c) Beam pattern of LABS at a single wavelength. (d) Beam pattern of wavelength tuning at a single emitter.

## S2. Design of staircase grating emitter

A staircase grating is proposed in our beam-steering device to increase the emission efficiency. We have compared the upward emission efficiency of staircase and normal grating structures based on 2D FDTD simulation. As shown in Fig. S2(a), with a parameter optimization based on particle swarm algorithm, the highest efficiency is found to be ~0.88 and ~0.49 at 1550 nm for 40-μm-long staircase and normal gratings, respectively.

Considering that a grating with long radiation length is required to obtain small beam divergence in the far field according to Eq. (6), there is a tradeoff between the grating length and emission efficiency during the grating design. Parameters of grating are shown in Fig. S2(b) including stage width $a_1$, $a_2$, $a_3$ and etching depth $h_1$, $h_2$. These parameters are restricted by three conditions, i.e., fabrication limitation, single order diffraction requirement and limited numerical aperture of lens, given by:

$$\frac{\lambda_0}{n_{eff} + 0.55 n_{sio_2}} < a_1 + a_2 + a_3 < \frac{\lambda_0}{n_{eff} - 0.55 n_{sio_2}} \quad \text{(S1)}$$

$$\frac{\lambda_0}{2n_{eff}} < a_1 + a_2 + a_3 < \frac{\lambda_0}{n_{eff}} \tag{S2}$$

$$a_x > 150nm, h_y > 30nm, (x=1,2,3, y=1,2) \tag{S3}$$

The width of the first stage $a_1$ is fixed as 0.2 μm to reduce the simulation complexity. The simulation results of upward emission efficiency $T_{up}$ and the effective grating length $L$ with different values of stage widths are shown in Figs. S3(a)-(d) under an etching depth of $h_1=h_2=40$ nm, 50 nm, 60 nm, and 70 nm, respectively, where the contour map with color fill represents upward emission efficiency, the contour with red lines illustrates the effective grating length, and the contour with white lines is the grating period. As expected, there exist opposite trends for upward emission efficiency and effective grating length, i.e., a high emission efficiency means a strong radiation strength along grating and leads to a short effective grating length. An effective grating length of 0.2 mm is required in our experiment, thus, the optimal selection lies in the marked optimal zone, as shown in Figs. S3(a)-(d).

Since the etching depth also affects the upward emission efficiency and effective grating length. The simulation results of emission efficiency and effective grating length under different etching depth and grating period are summarized in Fig. S4. The contour map with color shows an upward emission efficiency distribution and the solid red line illustrates the effective grating length (power decrease to 1/e) under different etching depth. Under the requirements of $T_{up}>60\%$ and $L>200$ μm in our experiment, we fabricated a staircase grating with a grating period of 0.85μm (a1=0.2 μm, a2=0.42μm, a3=0.23μm), etching depth $h_1=50$ nm, $h_2=50$ nm, $T_{up}=64\%$, and $L=0.2$ mm.

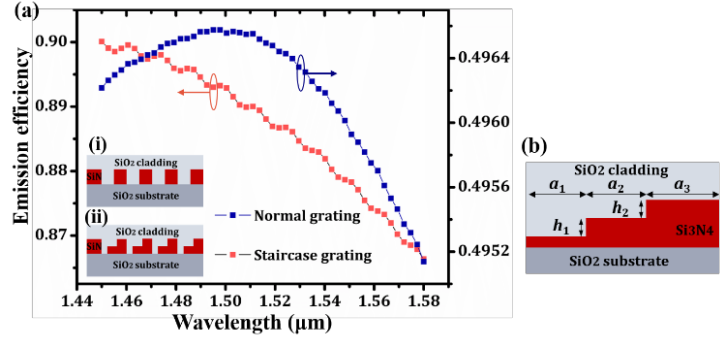

Fig. S2. (a) Comparison of maximum emission efficiency between normal and staircase gratings simulated by FDTD. Inset: cross section of (i) normal grating and (ii) staircase grating. (b) Structure of staircase grating in one grating period.

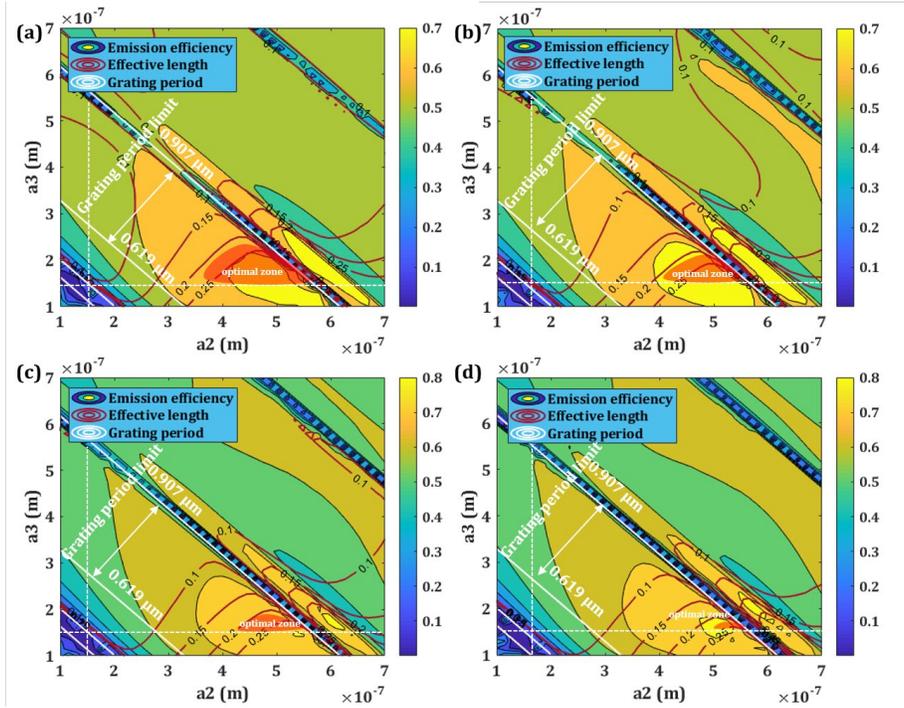

Fig. S3. Simulation emission efficiency and effective grating length of staircase grating versus grating period under etching depth of (a) 40 nm, (b) 50 nm, (c) 60 nm and (d) 70 nm.

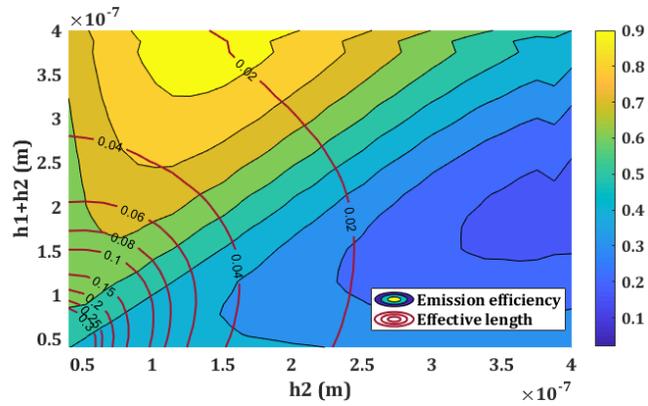

Fig. S4. Simulation results of upward emission efficiency versus etching depth $h_1$, $h_2$, radiation length $L$.

## S3. Emission efficiency of staircase grating.

The emission efficiency comparison between a normal single-etching grating and a staircase grating are performed experimentally. We directly analyze the emission beam brightness in the images captured by the infrared camera. Considering the response of the camera is not linear, we first use an 18.6-μm-long grating for calibration, as shown in Fig. S5. For each input power, the brightness value (0~65535, 16 digits) of the captured beam spot is recorded. Then the actual emitted power from the grating is calculated with the input power and grating insertion loss. The grating insertion loss is measured at a fixed input power by placing a large-core multi-mode fiber close to the grating to obtain the emitted power. The insertion loss of grating is then obtained and assumed to be constant as the input power is low. A fitting curve is also plotted in Fig. S5. The insets show some typically images of the beam spots.

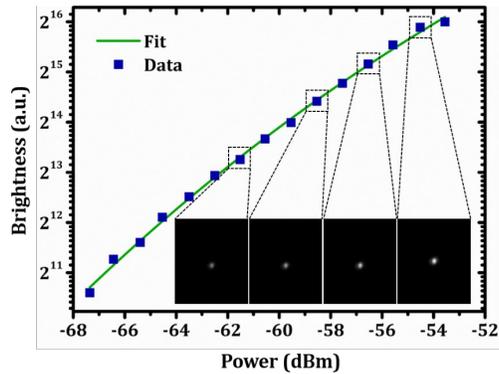

Fig. S5. Camera response calibration with a normal coupling grating. Inset: the grating beam patterns captured by infrared camera.

Figure S6 shows the captured near-field beam pattern of the staircase grating and a normal single-etching grating. The images are captured from different angles to search for the emission angle of the gratings. It is found that the emission angles are 6° for staircase grating and 18° for normal grating, respectively. Together with the calibration result mentioned above, the intensity curves for both gratings are obtained, as shown in Fig. 4(b). The relative power of the staircase grating is 1.23 times of normal grating.

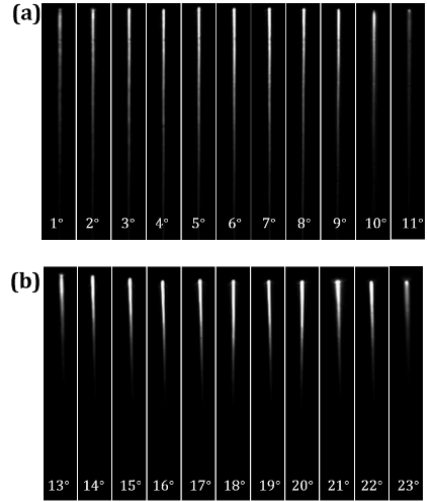

Fig. S6. Near-field beam patterns of (a) staircase grating and (b) normal grating versus detection angles of infrared camera.

## S4. Silicon-nitride thermo-optical switch

The optical switch is fabricated using an MZI structure with thermal tuning, where the heating resistors are titanium film deposited on silica cladding, and the wiring lines and bonding pads are aluminum. It is measured to have a switching time of 50 μs with an extinction ratio of 25 dB, and a power consumption of 120 mW/π as shown in Fig. S7.

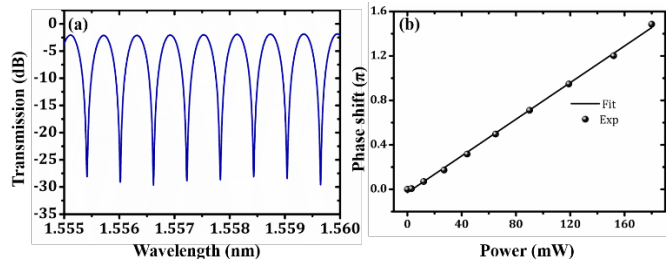

Fig. S7. (a) A typical transmission spectrum of a 1×2 MZI switch with unequal arms. (b) Measured phase shift at different electrical power.

## S5. Calculation of blind zone suppression

As the parameter blind zone suppression (BZS) is to evaluate the blind zone in LABS technology, we let BZS = '-' for OPA in [1].

For LABS in [2], the beam emission is achieved by edge emitters. Based on the layout in Fig. 1(b) in [2], the BZS is the ratio between the edge emitter width and the edge emitter pitch, which is estimated to be 10%~15%. With some methods of non-uniform waveguide design developed in OPA for half-wavelength emitter pitch, e.g., slightly different propagation constant in the waveguides [3], the BZS can be increased to ~50%. However, these methods only work for 1D edge emitters. Until today, no solution for 2D high-density on-chip emitter array has been experimentally demonstrated yet.

For LABS in [4], the BZS is equal to the ratio between the waveguide area under MEMS grating and the total MEMS area. In Fig. 2(b) in [4], the waveguide area under MEMS grating is estimated to be 1×8 μm$^2$ and the total MEMS area is estimated to be 60×60 μm$^2$. The BZS is therefore 8/3600=0.22%. If the minimum MEMS area can be reduced to 10×10 μm$^2$, the BZS is then equal to 8%. In the table, we input <10% as the estimated maximum value for this MEMS based LABS technology.

For our previous work with a SiO$_2$ chip and fiber array [5], the duty cycle is the ratio between the fiber core area and the area occupied by the fiber. For single mode fiber with 10 μm core and 125 μm cladding, the duty cycle is equal to 78.5 μm$^2$/125$^2$ μm$^2$=0.5%. If some multi-mode fiber is used, for example 100 μm core and 125 μm cladding, the BZS can be increased to ~50%.

For our current work, the BZS is directly calculated by the grating width divided by the grating pitch, that is 11 μm/(11+0.6) μm=95%, where 11 μm is the grating width, 0.6 μm is the minimum allowable grating gap.